\documentclass[aps,notitlepage,nofootinbib]{revtex4-2} 

\usepackage{graphicx} 
\usepackage{subcaption}  
\usepackage{amsmath}
\usepackage{xcolor,soul}

\begin{document}

\title{Spontaneous Leptogenesis in Type I Seesaw}

\author{Eung Jin Chun}
\affiliation{Korea Institute for Advanced Study, Seoul 130-722, Korea}

\author{Hyun Min Lee and  Jun-Ho Song}
\affiliation{Department of Physics, Chung-Ang University, Seoul 06974, Korea}

\begin{abstract}
Type-I seesaw models with a spontaneously broken $B-L$ symmetry provide a natural framework for spontaneous leptogenesis driven by a Majoron. The kinetic background of the Majoron acts as a CP-violating source, generating a lepton asymmetry both through the decay of right-handed neutrinos and through equilibration via inverse-decay processes. We construct the Boltzmann equations in a fully consistent manner, incorporating both effects, to enable a quantitative analysis. When the neutrino Yukawa coupling is large enough to maintain $B-L$ violating interactions in thermal equilibrium, the resulting asymmetry closely tracks its equilibrium value. In contrast, when this condition is not satisfied, a nontrivial interplay emerges between decay and inverse-decay dynamics, determined by the Yukawa coupling strength and the initial abundance of right-handed neutrinos.
\end{abstract}

\maketitle

\section{Introduction}

Spontaneous baryogenesis stands as one of the most compelling approaches for explaining the universe's matter-antimatter asymmetry. This mechanism, first proposed in \cite{CK87}, operates within a CPT-violating background created by the classical motion of a pseudo-Nambu-Goldstone boson (pNGB) that arises from the spontaneous breaking of a $U(1)$ symmetry. The core feature is that the kinetic motion of the pNGB acts as an ``effective chemical potential'' for $U(1)$-charged particles, and then it can subsequently translate into a chemical potential for baryon number ($B$) while $B$-violating interactions remain in thermal equilibrium. The resulting baryon asymmetry can be maintained at the observed level if these $B$-violating processes freeze out before the kinetic motion, assumed to originate in the early universe, significantly diminishes.

Recently, it has been recognized \cite{CH19} that this concept can be effectively implemented within the Peccei–Quinn framework, which addresses the strong CP problem through the anomaly coupling of the QCD axion (the pNGB of the spontaneously broken $U(1)_{PQ}$). In this setting, the axion-induced effective chemical potential drives the baryon (or, more precisely, $B+L$) asymmetry via electroweak sphaleron processes, which cease to be effective after the electroweak phase transition. A thorough investigation of this specific mechanism was provided in \cite{DEMY20}.

In a similar vein, spontaneous leptogenesis can be naturally realized in a widely studied scenario for explaining neutrino masses and mixing: the Type-I seesaw model extended by a spontaneously broken $U(1)_{B-L}$ symmetry \cite{CMP80}. Here, the kinetic motion of the corresponding pNGB, Majoron, provides the seed for the lepton asymmetry. This occurs while the $B-L$-violating neutrino Yukawa interactions remain in equilibrium at temperatures near the mass scale of the right-handed neutrinos \cite{CJ23,Wada24}. This should be contrasted with earlier studies on Majoron-driven spontaneous leptogenesis, which arises only after integrating out the heavy right‑handed neutrinos \cite{IK15,CP23}.  Various attempts to achieve leptogenesis in the context of the axion or axion-like particles, which do not couple to the $B-L$ current, have been extensively explored \cite{KSY14,CFGHH20,Berbig23,BCHP24,DMS24,CMSV24}. The present study can be extended to type-II seesaw models \cite{Berbig25} as well as the Peccei-Quinn framework in which the QCD axion is identified with the Majoron \cite{BPK24,Chun25,Chun26}. 

\medskip

In this context, an important observation was made in \cite{BMS17}: when a right-handed neutrino couples to a CPT-violating background, such as one provided by a constant torsion field strength, its decay can generate a lepton asymmetry even at tree-level, which distinguishes it from conventional thermal leptogenesis \cite{FY86}. This coupling is equivalent to the derivative interaction between the Majoron and fermions, and its influence on the decay kinematics must be rigorously incorporated into the analysis of spontaneous leptogenesis.

The purpose of this article is to provide a consistent and comprehensive analysis of baryon (or, more precisely, $B=L$) asymmetry generation within the framework of spontaneous leptogenesis in the Type-I seesaw model, accounting for the effects arising from both right-handed neutrino decays and equilibration processes. We anticipate that the final lepton asymmetry will be dictated by one of two regimes: its equilibrium value (when the neutrino Yukawa coupling is sufficiently large to maintain $B-L$ violation until the right-handed neutrino is non-relativistic), or the dynamic interplay between the decay and inverse-decay processes (in the opposite regime). 

To elucidate and quantify these essential features, we formulate the effective chemical potential for heavy Majorana fermions (as well as Dirac or massless fermions) and analyze its impact on right-handed neutrino decay. This allows us to construct the Boltzmann equations in the presence of a Majoron kinetic background, thereby establishing a robust framework for studying the evolution of the $B-L$ asymmetry.


\section{Spinors in a CPT Violating Background}

We begin with the leptonic part of the Standard Model (SM) Lagrangian extended by right-handed Majorana neutrinos $N_R=N_L^c$:
\begin{equation}
{\cal L}=y_{\nu} \bar{l}_L H N_R + {1\over2} y_N S \bar{N}_L N_R +h.c 
\end{equation}
where $l_L=(\nu_\alpha, \alpha)_L$  is the left-handed lepton doublet with $\alpha=e,\mu, \mbox{or } \tau$, $H=(H^0,H^-)$ is the Higgs doublet, and $S$ is a scalar field carrying $U(1)_{B-L}$ charge $x_{S}=1$. This assignment implies $x_{l_L}=x_{N_R}=-1/2$. After $B-L$ symmetry breaking, the scalar takes the form 
$$S\equiv {f_a \over \sqrt{2}} e^{i \theta}, \quad \theta\equiv {a \over f_a}$$ 
where the PNGB $a$ (Majoron) is assumed to acquire a small mass from a soft breaking term.  The right-handed neutrino then obtains a mass $m_N=y_N f_a/\sqrt{2}$. 
Removing the phase field $\theta$ through a $B-L$ rotation, $\psi\to e^{i x_\psi \theta}\psi$ where $x_\psi$ denotes the $B-L$ charge of a chiral fermion $\psi$,\footnote{Note that this defines our charge assignment $x_\psi ={1\over2} (B-L)_\psi$}  the Majoron couples to SM fermions only through derivative interactions:
\begin{equation} 
{\cal L}_\theta = -x_\psi \theta_\mu  \bar{\psi} {\gamma}^\mu \psi ,
\end{equation} 
with $\theta_\mu \equiv \partial_\mu \theta=(\dot\theta,\vec{0})$ treated as a kinetic background. Here, $\psi$ denotes any relevant chiral fermion, e.g., $\psi=N_R, l_L$, etc..
For clarity, we adopt the chiral (Weyl) representation of Dirac matrices: 
\begin{equation*}
\gamma^\mu =
\begin{pmatrix} 
0 & \sigma^\mu \\ 
\bar{\sigma}^\mu & 0 
\end{pmatrix}  ,  \quad
\sigma^\mu=(I, \vec{\sigma}), ~ 
\bar{\sigma}^\mu=(I, -\vec{\sigma}),
\end{equation*}
so that  a four-component spinor $\Psi$ decomposes as $\Psi=(\psi_L, \psi_R)$.  

\medskip

For a {\bf Dirac fermion}, $\psi\neq \psi^c$ (equivalently $\psi_L \neq {\psi_R}^{c}$), such as a charged lepton moving in the $\dot\theta$ background, the $u$-spinor of $\psi\sim u(p) e^{-ip\cdot x}$ satisfies
\begin{equation}\label{Duspinor}
  \begin{cases}
(p_\mu -x_\psi \theta_\mu) \bar{\sigma}^\mu u_L(p) =m_\psi u_R(p),  \\
(p_\mu -x_\psi \theta_\mu) \sigma^\mu u_R(p) =m_\psi u_L (p) . 
  \end{cases} 
\end{equation}
The corresponding  $v$-spinor equation follows from $\psi\sim v(p) e^{ip\cdot x}$ can be obtained by replacement: $p_\mu \to -p_\mu$ and $u(p)\to v(p)$. 
The resulting dispersion relation is 
\begin{equation} \label{Ddispersion}
    E=E_0 \pm x_\psi \dot \theta, ~~~ E_0\equiv \sqrt{\vec{p}^2+m_\psi^2},
\end{equation}
which differs for the particle ($u$) and antiparticle ($v$) solutions. Thus, $\dot\theta$ acts as an effective chemical potential, inducing an asymmetric thermal distribution between particles and antiparticles. 

Explicitly, the spinor solutions are 
\begin{equation} \label{Dspinors}
    \begin{cases}
        u_L^r(p)=\sqrt{p_- \cdot \sigma} \,\xi_r ,\\
        u_R^r(p)=\sqrt{p_- \cdot \bar{\sigma}}\, \xi_r ,
    \end{cases}\qquad
    \begin{cases}
      v_L^r(p)=\sqrt{p_+ \cdot \sigma}\, r\xi_{-r} , \\
        v_R^r(p)=\sqrt{p_+ \cdot \bar\sigma}\, (-r)\xi_{-r} ,
    \end{cases}
\end{equation}
where $p_{\mp,\mu}\equiv p_\mu \mp x_\psi \theta_\mu$ and $r=\pm1$ labels spin eigenstates.  Notably,  these spinors reduce to their standard forms when using the dispersion relation  above, since the $\dot\theta$-dependent shifts cancel out:  $p_{\mp \mu}=(E_0, \vec{p})$.

\medskip

The situation differs for a {\bf Majorana fermion}, $\psi_L={\psi_R}^c$, whose two chiral components carry opposite charges: $x_{\psi_L}=-x_{\psi_R}\equiv x_\psi$. The $u$-spinor now satisfies
\begin{equation}\label{uspinor}
  \begin{cases}
(p_\mu -x_\psi \theta_\mu) \bar{\sigma}^\mu u_L(p) =m_\psi u_R(p), \\
(p_\mu +x_\psi \theta_\mu) \sigma^\mu u_R(p) =m_\psi u_L (p) .
  \end{cases}
\end{equation}
The resulting dispersion relation is
\begin{equation} \label{Mdispersion}
E=\sqrt{m_\psi^2+ (|\vec{p}|-{\cal H} x_\psi \dot \theta)^2} ~~\overset{E_0 \gg |\dot\theta|}{\longrightarrow}~~ E_0 -{\cal H} {|\vec{p}| \over E_0} x_\psi \dot\theta ,
\end{equation}
where ${\cal H}\equiv \vec{p}\cdot \vec{\sigma}/|\vec{p}|=\pm1$ denotes helicity.  Thus, $\dot\theta$ acts as an effective chemical potential, but now its effect depends on helicity.  In the massless limit, where helicity aligns with chirality,  the Dirac and Majorana dispersion relation coincide. 
The corresponding spinors are now
\begin{equation} \label{Mspinors}
    \begin{cases}
        u_L^s(p)=\sqrt{p_+ \cdot \sigma} \,\xi_s \\
        u_R^s(p)=\sqrt{p_- \cdot \bar{\sigma}}\, \xi_s
    \end{cases}\qquad
    \begin{cases}
      v_L^s(p)=\sqrt{p_- \cdot \sigma}\, s\xi_{-s} \\
        v_R^s(p)=\sqrt{p_+ \cdot \bar\sigma}\, (-s)\xi_{-s}
    \end{cases}
\end{equation}
where $p_{\pm \mu} \equiv p_\mu \pm x_\psi \theta_\mu$ and $s=\pm1$ denotes helicity. Such a helicity-dependent shift  induces an effective particle-antiparticle asymmetry in the decay of  right-handed neutrinos. 

\section{Boltzmann Equations in Spontaneous Leptogenesis}

Using the spinor solutions (\ref{Dspinors}) and (\ref{Mspinors}) for leptons and right-handed neutrinos, respectively, one can straightforwardly compute the squared amplitudes, $|{\cal M}|_s^2$ and $|{\cal \bar{M}}|_s^2$, for the decay of a right-handed neutrino to leptons and antileptons, 
$$N_s \to l \bar{H}, \quad N_s \to \bar{l} H,$$  
for each helicity state $s=\pm1$. Tl leading order in $\dot\theta$, we obtain
\begin{equation} \label{squaredM}
    \begin{split}
        |{\cal M}|_s^2 = |y_\nu|^2 \left(E_l^0-s |\vec{p_l}| \cos\vartheta\right)\left(E_N^0+s |\vec{p}_N| -{\dot\theta\over 2}(1+ s{|\vec{p}_N| \over E_N^0})\right) \\
          |{\cal \bar{M}}|_s^2 =  |y_\nu|^2 \left(E_{\bar l}^0+s |\vec{p}_{\bar l}| \cos\vartheta\right)\left(E_N^0-s |\vec{p}_N| +{\dot\theta\over 2}(1- s{|\vec{p}_N| \over E_N^0})\right)  
    \end{split}
\end{equation}
where $\cos{\vartheta}=\vec{p}_l \cdot \vec{p}_N/|\vec{p}_l||\vec{p}_N|$.
This expression exhibits several consistency checks.
For the decays $N_{\pm} \to l^- \bar{H}$, only the $s= +1$ helicity  state contributes in the relativistic limit, corresponding to right-handed chirality. In this limit, helicity conservation requires the charged lepton momentum to be antiparallel to that of the decaying right-handed neutrino ($\vartheta=\pi$).
Integrating over the phase space in the static limit $\vec{p}_N\to 0$, one finds that the total decay rates satisfy 
\begin{equation} \label{Epsilon}
{\Gamma(l)-\Gamma(\bar{l}) \over \Gamma(l)+\Gamma(\bar{l})}=- {\dot\theta \over 2 m_N},
\end{equation}
showing explicitly that the kinetic Majoron background induces a lepton asymmetry even in the absence of conventional CP violation from the Yukawa couplings.

\medskip

Using the squared amplitudes (\ref{squaredM}) together with the dispersion relations (\ref{Ddispersion}) and (\ref{Mdispersion}) in the phase-space integrals for a massless lepton and a heavy right-handed neutrino, we derive the Boltzmann equations for 
$$Y_N\equiv Y_{N_+}+Y_{N_-},\quad Y_{\Delta N}\equiv Y_{N_+}-Y_{N_-}, \quad Y_{\Delta f}\equiv Y_{f}-Y_{\bar f},$$
where $Y_{f}=n_{f}/s$ is the number density of a field $f$ ($f=N_R, l_L, H$ in our system)  normalized by the entropy density $s={2\pi^2 g_* T^3/ 45}$.
It is convenient to work with the  $B-L$ asymmetry $Y_{B-L}$ which is conserved in the Standard Model. 
Keeping terms up to first order in $\dot\theta$ and imposing chemical equilibrium among Standard Model interactions, we find that the equations for $Y_{\Delta N}$ and $Y_{B-L}$ decouple from each other and the relevant Boltzmann equations become
\begin{equation}\label{Boltz}
    \begin{split}
        {d Y_N \over d z} &= -z K \gamma_D \left(Y_N-Y_N^{\rm eq}\right)  \\
        {d Y_{B-L} \over d z} &= z K \gamma_M \varepsilon_1 \left(Y_N-Y_N^{\rm eq}\right) -z K \gamma_{ID} \left( Y_{B-L}+{y_\theta \varepsilon_1 \over z^2}) \right) 
    \end{split}
\end{equation}
where $z\equiv {m_N / T}$, $K\equiv {\Gamma_N / H_1}$ is the ratio of the decay rate $\Gamma_N\equiv {y_\nu^2 m_N/8\pi}$ and the Hubble parameter $H_1\equiv H(z=1)$. The quantity $y_\theta \varepsilon_1$ is a constant parameterizing the $B-L$ number stored in the Majoron background $\dot\theta$ at $T=m_N$:
\begin{equation}
 y_\theta \equiv {1\over3} {45 \over 2\pi^2}{1\over g_*}, ~~~ \varepsilon_1 \equiv c_{l} {\dot\theta_1 \over  2m_N},
\end{equation}
where $\dot\theta_1=\dot\theta(z=1)$. The coefficient  $c_l$ depends on the flavor structure and chemical equilibrium conditions. For flavor-independent interactions, one finds $c_l=-79/11$, while flavor-dependent effects modify this by an order-one factor \cite{DNN08}. We present our results for $Y_{B-L}$ normalized to $y_{\theta}\varepsilon_1$. 

\begin{figure}[!ht]
    \centering
    \includegraphics[width=0.5\linewidth]{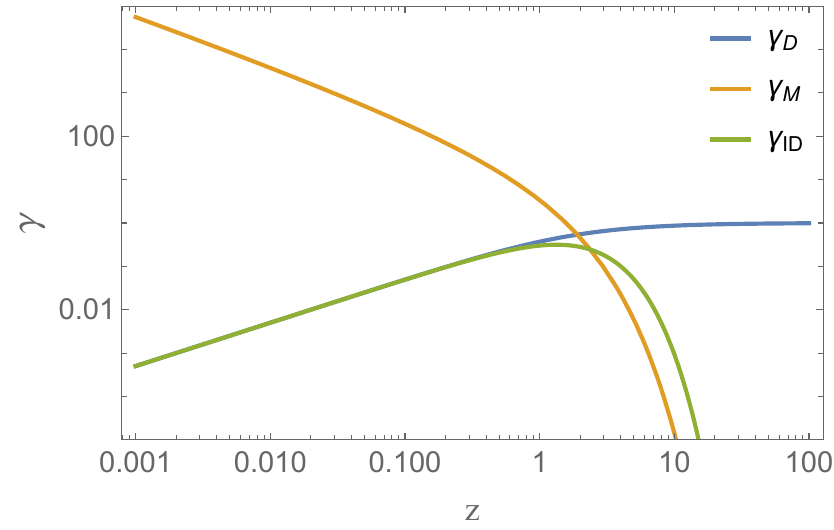}
    \caption{Evolution of $\gamma_{D,M, ID}$ in terms of $z=m_N/T$.}
    \label{fig:gamma}
\end{figure}

The functions $\gamma_{D, ID,M}$ used in (\ref{Boltz}) are defined as 
\begin{equation} \label{gammas}
    \begin{split}
        \gamma_D &= {z^2 K_1(z) \over z^2 K_2(z)}, \qquad \gamma_{ID}={1\over2} z^2 K_2(z), \\
        \gamma_M &= {G_M(z) \over z^2 K_2(z)}, ~~~~
        G_M(z)\equiv \int d|\vec{p}_N| e^{-{E_N^0 \over T}} {2|\vec{p}_N|\over E_N^0} 
        \ln\left( {E_N^0+|\vec{p}_N| \over E_N^0 -|\vec{p}_N|}\right)^2,
    \end{split}
\end{equation}
where $E_N^0=\sqrt{m_N^2+|\vec{p}_N|^2}$.

Before solving the Boltzmann equations, it is instructive to examine the behavior of the functions  $\gamma_{D},\gamma_{ID}$ and $\gamma_M$, shown in Figure 1.   The first two are familiar from conventional leptogenesis \cite{DNN08}, whereas $\gamma_M$ appears here due to the helicity-dependent  chemical potential of $N_R$. In standard leptogenesis, the inverse-decay washout term in (\ref{Boltz}) can remain efficient even at $T<m_N$ for $K>1$, despite the Boltzmann suppression of $\gamma_{ID}$. In spontaneous leptogenesis, this term plays an additional role: it couples to the CP-violating source $Y_\theta \propto \dot\theta$, thereby maintaining $Y_{B-L} = -c_l Y_\theta$ as long as inverse decays remain in equilibrium \cite{CJ23}.  
The new feature explored in this work is the decay term proportional to $\dot\theta$ whose efficiency is governed by $\gamma_M$. Since the helicity chemical potential decreases as $T/m_N$ (see (\ref{Mdispersion})), $\gamma_M$ falls more rapidly than $\gamma_{ID}$ at low temperature, but is more important for $T>m_N$ approaching the massless limit.  Consequently, the decay term is expected to dominate the generation of lepton asymmetry for $K<1$, provided that the heavy neutrino abundance $Y_N$ does not closely track its equilibrium value $Y_N^{\rm eq}$.

\medskip

\begin{figure}[!ht]
    \centering
   \includegraphics[width=0.32\linewidth]{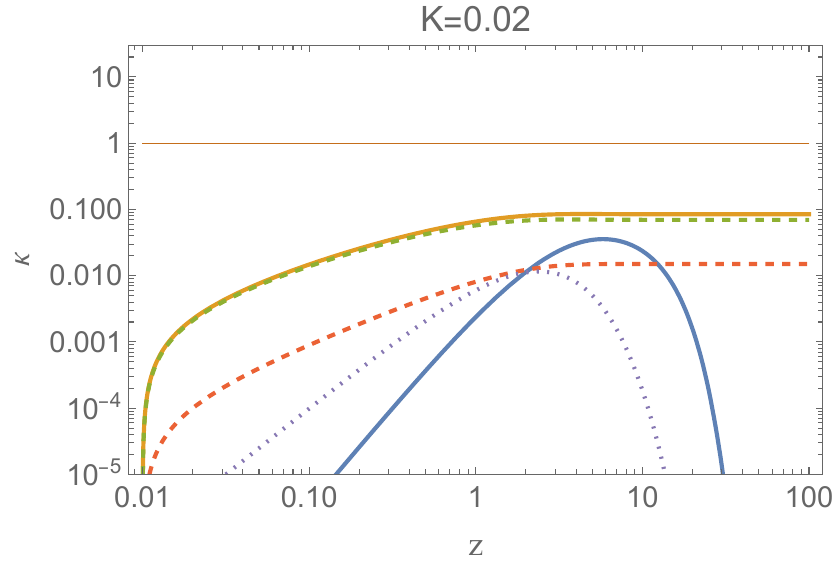}
   \includegraphics[width=0.32\linewidth]{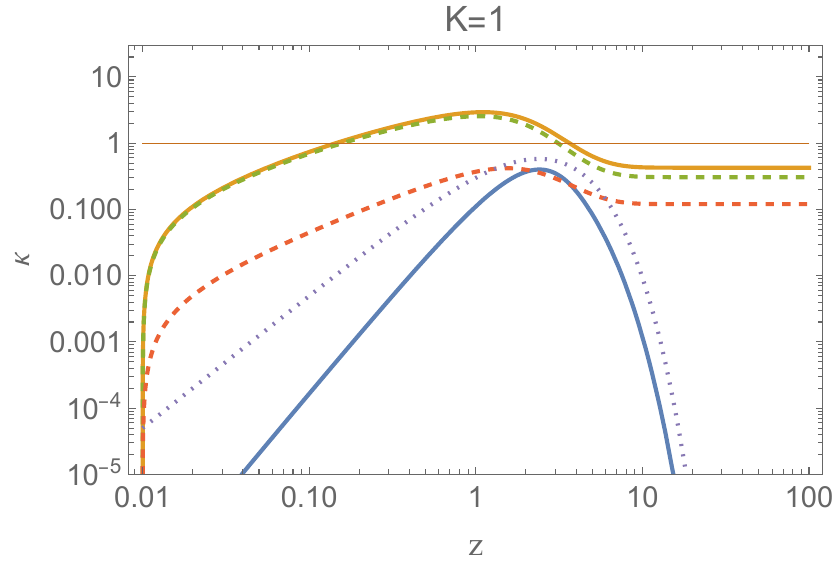}
    \includegraphics[width=0.32\linewidth]{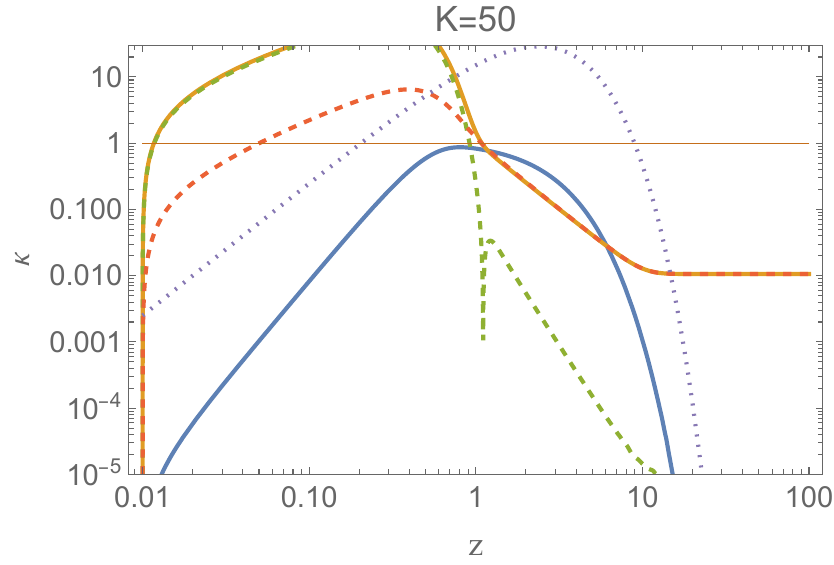}\\
\includegraphics[width=0.32\textwidth]{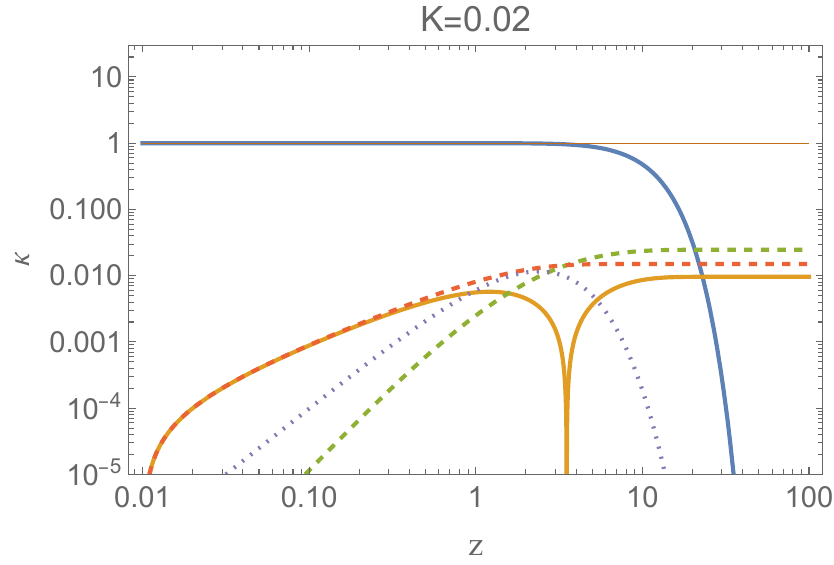}
\includegraphics[width=0.32\textwidth]{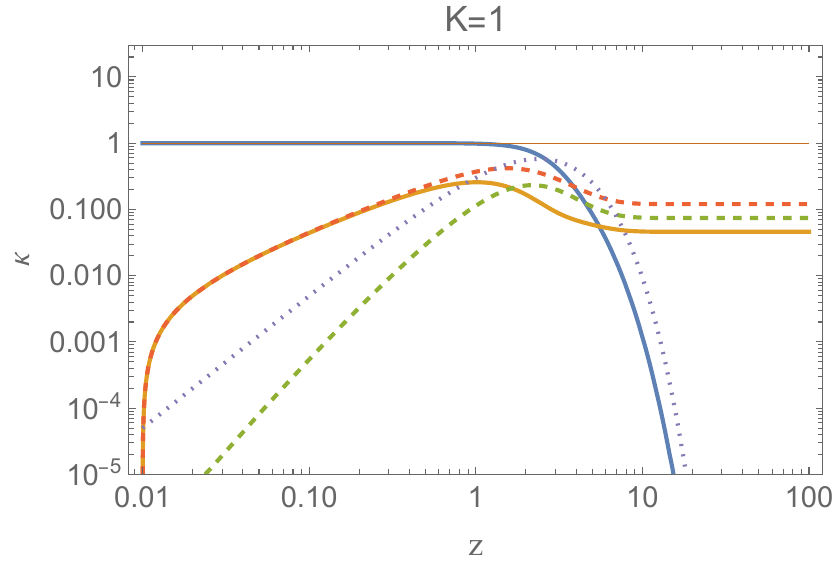}
\includegraphics[width=0.32\textwidth]{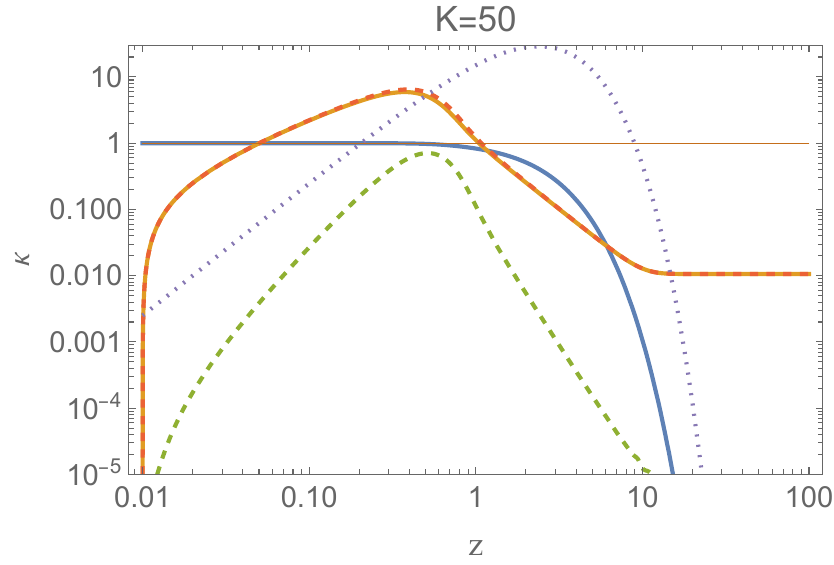}
\caption[short]{\raggedright Evolution of the efficiency $\kappa \equiv Y_{B-L}/y_\theta \varepsilon_1 $ for various values of $K$. The dashed green (red)  lines show the contribution from the decay (inverse-decay) processes alone, while the solid orange curves represent  their combined effect. The solid blue curves display the ratio $Y_N/Y_N^{\rm eq}$, and  the dotted curves indicate the inverse-decay rate normalized to the Hubble rate. The upper and lower panels correspond to the initial conditions $Y_N(0)=0$ and $Y_N(0)=Y_N^{\rm eq}$, respectively.}
    \label{fig:YL(z)}
\end{figure}

Figure 2 illustrates  the evolution of the normalized $B-L$ asymmetry $\kappa \equiv Y_{B-L}/y_\theta \varepsilon_1 $ measuring the efficiency of the spontaneous leptogenesis, and of $Y_N/Y_N^{\rm eq}$ for various values of $K$. 
The dashed green and red curves show the contributions from decays and inverse decays, respectively, while the dotted curves represent the inverse-decay rate normalized to the Hubble expansion rate.
The upper panels correspond to the case of vanishing initial abundance, $Y_N(0)=0$, while the lower panels assume thermal abundance, $Y_N(0)=Y_N^{\rm eq}$. For  $Y_N(0)=0$, the final asymmetry is dominated by the decay term for $K\ll1$, by inverse decays for $K\gg1$, and receives comparable contributions from both when $K\sim 1$. 
For  $Y_N(0)=Y_N^{\rm eq}$. the decay and inverse-decay contributions have opposite signs and (partially) cancel for $K\lesssim1$, leading to suppressed asymmetry. For $K\gg1$, inverse decays dominate in both cases, producing identical final asymmetries. 

\medskip

\begin{figure}[!ht]
    \centering
\includegraphics[width=0.45\linewidth]{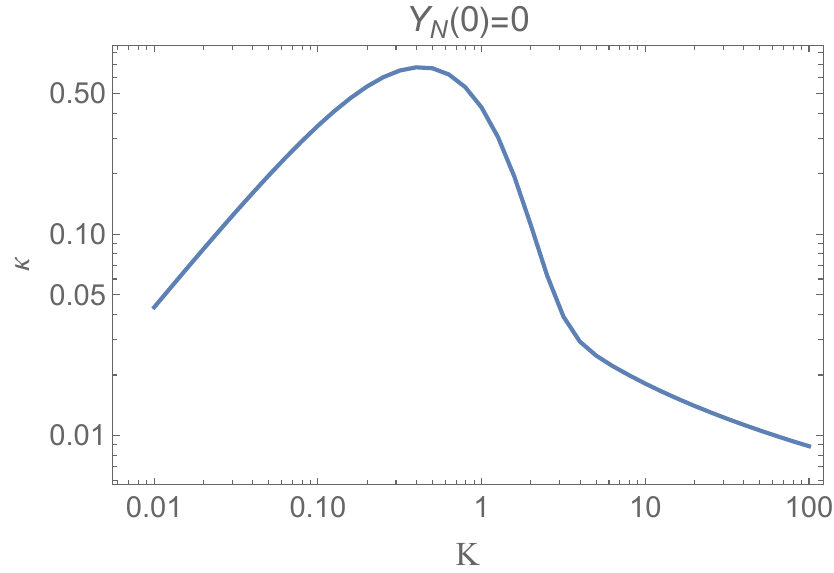} 
\hspace{1.5em}
\includegraphics[width=0.45\textwidth]{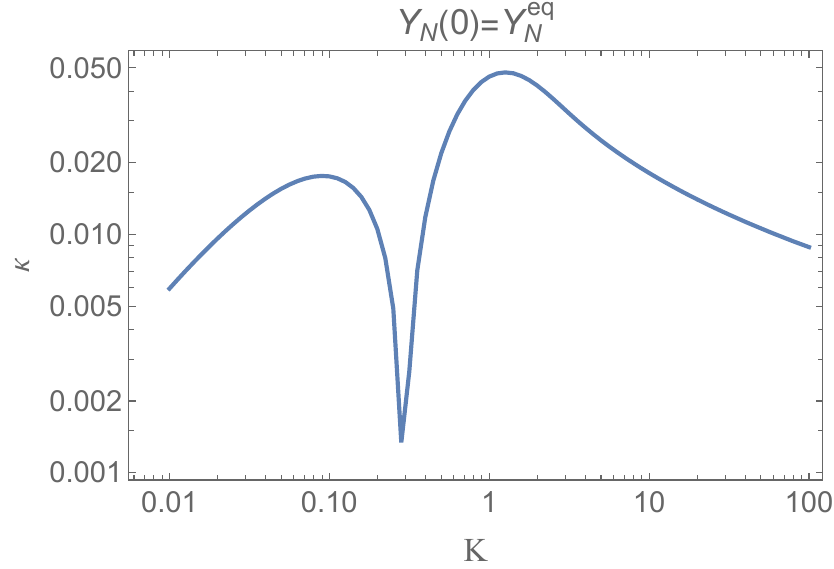}
    \caption[short]{\raggedright The efficiency $\kappa$  as a function of $K$ for two different initial conditions of $Y_N(0)=0$ and $Y_N^{\rm eq}$.}
    \label{fig:YL_K}
\end{figure}

Figure 3 summarizes the results by showing the efficiency $\kappa$ of the final  
$Y_{B-L}$  as a function of $K$ for both initial conditions. For thermal initial abundance, the cancellation between decay and inverse-decay contributions becomes exact near $z \simeq 0.3$, suppressing the asymmetry for $K<1$.
For $K\gtrsim 4$, both results converge, reflecting the equilibration driven by inverse decays.

\medskip

We close with a few remarks on the phenomenological implications of the results obtained above.
In our framework, the resulting $B-L$ asymmetry  depends on two input parameters: the decay rate of $N$ through the quantity $K=\Gamma_D/H\big|_{T=m_N}$ and the initial kinetic misalignment $\dot\theta$ through the CP asymmetric quantity $\varepsilon_1=\dot\theta/2T\big|_{T=m_N}$. Given these, the $B-L$ symmetry breaking scale $f_a$ and the right-handed neutrino mass $m_N$ (or equivalently, the corresponding Yukawa coupling $y_N$) remain free parameters. Consequently, low-scale leptogenesis can be realized provided that an appropriate initial kinetic misalignment is arranged.  This stands in contrast to standard thermal leptogenesis, which operates at a high scale $m_N\gtrsim 10^9$ GeV, barring resonance conditions \cite{DI02}. 
It is instructive to compare the efficiency of spontaneous leptogenesis with that of standard thermal leptogenesis. Since $\varepsilon_1$ appearing in the Boltzmann equation for $Y_{B-L}$ in (\ref{Boltz}) plays the same role as the usual CP-asymmetric quantity, the efficiency factor $\kappa$ shown in Fig.~3 can be directly compared with that in Fig.~6 of \cite{BP04}. For $K\gtrsim 4$, where the efficiency is insensitive to the initial abundance, spontaneous leptogenesis becomes less efficient at larger $K$. For $K\lesssim 1$, our efficiency factor is higher by orders of magnitude in the case $Y_N(0)=0$.  The situation is reversed for $Y_N(0)=Y_N^{\rm eq}$, where the efficiency of standard thermal leptogenesis is maximal.

Let us note that coherent oscillation of the Majoron begins once the kinetic energy associated with the initial velocity $\dot\theta$ falls below the potential barrier set by a tiny Majoron mass $m_a$, introduced through a soft breaking of the $B-L$ symmetry \cite{CHH19}. This transition occurs at the moment when ${\dot\theta} \approx m_a$, and must take place after spontaneous leptogenesis has completed. This requirement imposes the condition $\dot\theta_1 \gg m_a$, where $\dot\theta_1 = 2 m_N Y_{B-L}/(\kappa c_l y_\theta)$, or equivalently, $m_a/{\rm eV} \ll (m_N/{\rm TeV})/\kappa $. 
The coherent oscillation of the Majoron is a compelling dark matter candidate,
With an appropriate choice of parameters, this framework thus enables Majoron cogenesis, simultaneously accounting for both the baryon asymmetry and the dark matter abundance \cite{CJ23}.

\section{Conclusion}

Spontaneous leptogenesis offers a compelling framework for explaining the observed matter-antimatter asymmetry of the Universe. This mechanism arises naturally in Type-I seesaw models with a spontaneously broken $B-L$ symmetry, where the kinetic background $\dot\theta$ of the associated Majoron provides the source of CP violation. In this setup, the net asymmetry is generated through two primary channels: the decay of right-handed neutrinos ($N_R$) and the equilibration of $B-L$ violating processes mediated by the neutrino Yukawa interactions. By consistently incorporating both effects, we formulate and solve the Boltzmann equations. In these equations, $\dot\theta$ plays a dual role: it induces a helicity chemical potential for $N_R$, producing asymmetric decay rates, while simultaneously acting as the seed for the $B-L$ violating equilibrium state via the inverse-decay process.

This resulting dynamics is analyzed in detail, and the main features are summarized in Figure 3. Regardless of the initial conditions--vanishing and equilibrium initial abundance, $Y_N(0)=0$ and  $Y_N(0)=Y_N^{\text{eq}}$--the final asymmetry approaches the equilibrium value whenever inverse decays remain efficient, corresponding to sufficiently large Yukawa couplings ($K\gtrsim 4$). In the intermediate regime ($K\sim 1$), the contributions from decay and inverse-decay become comparable, but their interplay depends sensitively on the initial condition: they add constructively for $Y_N(0)=0$ and destructively for $Y_N(0)=Y_N^{\text{eq}}$. Notably, the cancellation can be nearly complete around $K\simeq 0.3$, yielding a highly suppressed final asymmetry.

We expect that these results are broadly applicable and can be readily extended to a wide class of neutrino mass models,

\section*{Acknowledgments}

HML and JHS are supported in part by Basic Science Research Program through the National
Research Foundation of Korea (NRF) funded by the Ministry of Education, Science and
Technology (NRF-2022R1A2C2003567).

\end{document}